\documentclass[twocolumn,showpacs,preprintnumbers,amsmath,amssymb]{revtex4-1}
\usepackage{graphicx}
\usepackage{textcomp}

\begin{document}
%\draft
\title{Talbot self-imaging in $\mathcal{PT}$-symmetric complex crystals}
  \normalsize
\author{Stefano Longhi}
\address{Dipartimento di Fisica, Politecnico di Milano and Istituto di Fotonica e Nanotecnologie del Consiglio Nazionale delle Ricerche, Piazza L. da Vinci
32, I-20133 Milano, Italy}

%\date{.}
%
\bigskip
\begin{abstract}
The Talbot effect, i.e.  the self-imaging property of a periodic wave in near-field diffraction, is a remarkable interference phenomenon in paraxial systems with continuous translational invariance. In crystals, i.e. systems with discrete translational invariance, self-imaging has been regarded so far as a seldom effect, restricted to special sets of initial field distributions. Here it is shown that  in a class of gapless $\mathcal{PT}$-symmetric complex crystals at the symmetry breaking threshold Talbot revivals can arise for almost any initial periodic wave distribution which is commensurate with the lattice period. A possible experimental realization of commensurate Talbot self-imaging for light pulses in complex 'temporal' crystals, realized in an optical dispersive fiber loop with amplitude and phase modulators, is briefly discussed.
  \noindent
\end{abstract}

\pacs{42.25.Fx, 11.30.Er, 42.25.Bs}

%42.25.Fx 	Diffraction and scattering

\maketitle

\section{Introduction}
The Talbot effect \cite{Talbot,Rayleigh}, i.e. the self-imaging property of a periodic wave in near field diffraction,
is one of the most striking phenomena of wave physics \cite{review1,review2,review3}. Originally 
discovered for light waves \cite{Talbot,Rayleigh}, it was 
 later extended to matter waves \cite{matter1,matter2} and related to the wider class of revival phenomena in optics and quantum physics \cite{rev1,rev2}.
Nowadays the Talbot effect finds important applications in  
optical testing and
metrology, photolithography, spectrometry, 
and microscopy \cite{review2,review3,vari}.  Recently, several conceptual extensions of the Talbot effect
have been introduced, including revivals of nonclassical light \cite{quantumTalbot} (quantum Talbot effect), 
self-imaging phenomena in nonlinear systems \cite{nonTalbot} (nonlinear Talbot effect), sub-wavelength non-paraxial self-imaging in plasmonics and metamaterials \cite{cazz1} (super Talbot effect), and self-imaging in far-field diffraction patterns \cite{angular} (angular Talbot effect).
In Ref.\cite{discTalbot}, the concept of discrete Talbot effect, i.e. self-imaging in a lattice, was 
suggested and experimentally demonstrated for light beams in waveguide arrays. Unlike in the free space, where the
dispersion relation of the paraxial (Schr\"{o}dinger) wave
equation is parabolic, the discrete translational invariance of the lattice introduces allowed energy bands separated by energy gaps, making self-imaging  observable  for a  very limited set of periodicities of the input field \cite{discTalbot}. In Ref.\cite{PTTalbot} the discrete Talbot effect was extended to non-Hermitian lattices with parity-time ($\mathcal{PT}$) symmetry, however similar restrictions were found. Such results would suggest that self-imaging phenomena in systems with discrete translational invariance are unlikely or restricted to very special input field distributions. \par
In this work we shed new light into the physics of the Talbot effect in periodic lattices and show that, contrary to such a previous belief, complex crystals possessing $\mathcal{PT}$ symmetry can show self-images for rather arbitrary periodic field distributions with a spatial period $L$ which is {\it commensurate} with the lattice period $a$, i.e. 
 $L/a=N/M$ where $M$ and $N$ are arbitrary relatively prime numbers and $N$ odd. Unlike the ordinary Talbot effect in free space, where the Talbot revival period $z_T$ is finite for any period $L$ of the incident wave pattern, in complex crystals the revival period $z_T$ scales like $ML$ and thus diverges as $L$ becomes incommensurate with the lattice period $a$.\\
 The paper is organized as follows. In Sec.II we present a few general properties of Bragg diffraction and self-imaging phenomena in periodic potentials, and demonstrate the possibility to observe commensurate Talbot self-imaging in gapless complex crystals as opposed to approximate revivals in gapped potentials. In Sec. III  we describe in details self-imaging effects in a class of gapless complex crystals synthesized by supersymmetric transformations. A possible experimental realization of commensurate Talbot self-imaging for light pulses in complex 'temporal' crystals, realized in an optical dispersive fiber loop with amplitude and phase modulators, is presented in Sec.IV, whereas the main conclusions are outlined in Sec.V. Finally, three Appendices provide some technical and mathematical details of the study.
 
\section{Bragg diffraction and self-imaging in periodic potentials: general results} 
Let us consider Bragg scattering of optical or matter waves from a complex crystal [see Fig.1(a)]. In the paraxial approximation wave propagation is described by  the Schr\"{o}dinger-like wave equation \cite{Bragg1,Bragg2,Bragg3}
\begin{equation}
i \frac{\partial \psi}{\partial z} =-\frac{\partial^2 \psi}{\partial x^2}+V(x) \psi(x,z) \equiv \hat{H} \psi(x,z)
\end{equation}
where $z$ and $x$ are normalized paraxial propagation distance and spatial transverse coordinate, respectively, and $V(x)$ is the periodic optical potential with lattice period $a$. In a complex crystal, the optical potential $V(x)$ has a non-vanishing imaginary part, and the corresponding Hamiltonian $\hat{H}$ is thus non-Hermitian. Of particular relevance are complex crystals possessing $\mathcal{PT}$ symmetry $V(-x)=V^*(x)$ \cite{Bender}, which show an entire real energy spectrum in the unbroken $\mathcal{PT}$ phase \cite{Bender,Bender1,Bender2,Longhi,Jones}.  Complex potentials have been experimentally realized for matter waves \cite{Bragg1,Oberthaler} and optical systems \cite{uffa1,uffa2,uffa3,uffa4}. Compared
to ordinary crystals, wave transport in complex crystals exhibits some unique
properties, such as violation of the Friedel's law of Bragg
scattering \cite{Oberthaler,Bragg3}, double refraction and nonreciprocal diffraction \cite{Bragg2}, 
and unidirectional invisibility \cite{uffa3,uffa4,invisi}. Here we show that, in addition to such unusual properties and contrary to Hermitian lattices, $\mathcal{PT}$ complex crystals may allow for self-imaging of a wide set of input field distributions. Let us first recall that for free-space propagation, i.e. for  $V(x)=0$, an initial periodic wave $\psi(x,0)$ with spatial period $\mathcal{L}$, i.e. $\psi(x+\mathcal{L},0)=\psi(x,0)$, reproduces itself at propagation distances $z=nz_T$ ($n=1,2,3,...$), where $z_T= \mathcal{L}^2/(2 \pi)$ is the revival (Talbot) period. Moreover, at $z=(2n+1)z_T/2$ one has $\psi(x,z)=\psi(x+ \mathcal{L}/2,0)$, i.e. the same pattern recurs at odd multiplies of $z_T/2$ but shifted in space by half the spatial
periodicity. Self-imaging stems from the special parabolic dispersion relation $E=p^2$ of the energy spectrum of $\hat{H}=-\partial^{2}_{x}$ in free space for plane waves $\psi \sim \exp(ipx)$ with transverse momentum $p$. In the presence of a periodic potential $V(x)$, the  continuous translational invariance is broken and the energy spectrum comprises a sequence of allowed Floquet-Bloch bands separated by forbidden energy gaps \cite{Am}, namely $E=E_{\alpha}(q)$, where
$q$ is the Bloch wave number that varies in the
first Brillouin zone ($-\pi / a \leq q < \pi /a$ ) and $\alpha=0,1,2,3,...$ is the band index. 
Deviations of the dispersion curves $E_{\alpha}(q)$ from a parabolic law make it self-imaging phenomena unlikely in periodic potentials. For instance, as shown in Ref.\cite{discTalbot} for a single tight-binding lattice band, $E_{\alpha}(q) \propto \cos(q  a)$, the Talbot effect is possible only for a special set of initial field distributions with spatial period $\mathcal{L}=a,2a,3a,4a$ and $6a$. More severe restrictions have been found in a two-band lattice model \cite{PTTalbot}. Since for real periodic potentials there is at least one energy gap and the energy dispersion relations always deviate from a parabolic law (at least at low energies),  the Talbot effect is unlikely or subjected to severe restrictions as in the previously mentioned models. A remarkable property of complex crystals is to show no gaps in their energy spectrum, reproducing the parabolic dispersion relation of free space in spite of the discrete translational invariance of the system [see Fig.1(b)]. A gapless periodic potential of this kind is given, for example, by $V(x)=V_0 \exp( 2 \pi i x/a)$ \cite{Bender1,Longhi,Jones}. The gapless spectrum of this potential enables to observe such intriguing phenomena as unidirectional Bloch oscillations \cite{Bragg3} and unidirectional invisibility \cite{uffa3,uffa4,invisi}, however as shown in the following the Talbot effect is generally prevented here because of the existence of spectral singularities  \cite{Longhi,spectral} at both $q=0$ and $q=-\pi/a$. A natural question arises whether Talbot self imaging can be realized (and under which constraints) in any gapless complex crystal. Let $V(x)$ be a complex periodic potential of period $a$ with a gapless real energy spectrum $E \in [0, + \infty)$ and with parabolic dispersion relations $E_{\alpha}(q)=(2 \pi \beta_\alpha / a -|q|)^2$ of lattice bands, where $\alpha=0,1,2,3,...$ is the band index and $\beta_{\alpha}=0,1,-1,2,-2,3,-3,...$ for $\alpha=0,1,2,3,4,5,6,...$; see Fig.1(b). Then the following theorems hold (see Appendix A)\\
 \begin{figure}
\includegraphics[scale=0.4]{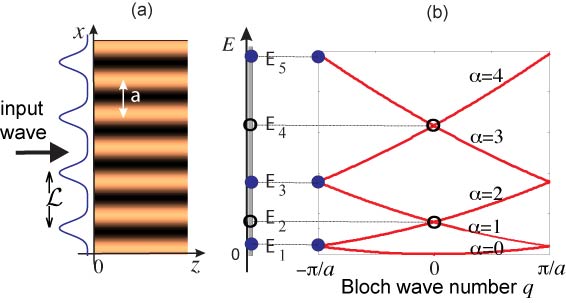}
\caption{(Color online) (a) Schematic of Bragg scattering and self-imaging in a complex crystal. The input wave at $z=0$ is periodic with a spatial period $\mathcal{L}$ which is commensurate with the lattice period $a$. (b) Band structure of a gapless crystal with a dispersion relation that reproduces the parabolic dispersion of free space propagation. Spectral singularities at energies $E_n=(n \pi/a)^2$may arise at either the band center $q=0$ (open circles, $n$ even) or at the band edge $q=-\pi/a$ (solid circles, $n$ odd). For the complex sinusoidal potential $V(x)=V_0 \exp(2 \pi i x/a)$ all the energies $E_n$ ($n=1,2,3,...$) are spectral singularities \cite{Jones}; for the potential (4) there is only one spectral singularity at $E=E_1$, whereas for the potential (5) there are two spectral singularities at energies $E_1$ and $E_3$. In all cases the band dispersion diagram is the same and obtained by folding, inside the first Brillouin zone, the parabolic dispersion curve of the free-particle Hamiltonian.}
\end{figure}
 {\it Theorem I.} Provided that the energies $E_{\alpha}(q=0)=0,(2 \pi/a)^2, (4 \pi/a)^2$, ... are not spectral singularities of $\hat{H}$, for an arbitrary initial field distribution $\psi(x,0)$ which is periodic in $x$ with a spatial period ${\mathcal L}=(N/M)a$ commensurate with the lattice period $a$, where $N$,$M$ are relatively prime numbers and $N$ is odd, then (i)  the propagated field $\psi(x,z)$ is periodic in $x$ with period $L=M \mathcal{L}=Na$; (ii)
 $\psi(x,z=n z_T)=\psi(x,0)$ for any $n=0,1,2,3,...$, where the revival (Talbot) period $z_T$ is given by 
\begin{equation}
z_T=\frac{N^2 a^2}{2 \pi}=\frac{M^2 \mathcal{L}^2}{2 \pi}.
\end{equation}
 {\it Theorem II.} Let $\psi(x,0)=f(x) \exp(2 \pi i p x/a)$, 
  where 
 $f(x+\mathcal{L})=f(x)$ is periodic with period $\mathcal{L}=(N/M)a$ commensurate with the lattice period $a$, $1/p$ is an integer number which is a prime number with respect to $N$, and $2Np$ is not an integer number. Then $\psi(x,nz_T)=\psi(x,0)$ with $z_T=(N^2 a^2)/(2 \pi p^2)$.\par

Theorem I basically states that Talbot self-imaging is possible for commensurate periodic field distributions in  gapless crystals that do not show spectral singularities at $q=0$, i.e. in none of the energies $E_2$, $E_4$, $E_6$, ... of Fig.1(b). Theorem II provides a sufficient condition for self-imaging in gapless crystals for a subset of initial field distributions $\psi(x,0)$, regardless of the existence of spectral singularities. Physically, such field distributions correspond to a periodic field $f(x)$ commensurate with the lattice period, which excites the crystal tilted at a certain angle that depends on the parameter $p$ ($p=1/2$ corresponds to the Bragg angle). Examples of complex crystals satisfying theorem I will be presented in Sec.III. Note that $z_T$ is $M^2$ times larger than the Talbot distance $\mathcal{L}^2/(2 \pi)$ in free space for the same period $\mathcal{L}$ of the initial field distribution. In particular, as $\mathcal{L}$ becomes incommensurate with the lattice period $a$, i.e. in the limit $N,M \rightarrow \infty$, according to Eq.(2) the Talbot revival distance $z_T$ diverges. This is a very distinct features as compared to the ordinary Talbot effect in free space, where the Talbot distance remains finite. Moreover, in complex crystals spatially-shifted self-images at odd multiplies of $z_T/2$ are not observed, as discussed in the examples of Sec.III.\par Hermitian lattices are not gapless, and thus self-imaging is unlikely. 
Nevertheless, the following theorem can be  stated (see Appendix A):\\
{\it Theorem III.}  Let $V(x)$ be a real periodic potential of period $a$. For a given initial field distribution $\psi(x,0)$ which is periodic with a spatial period ${\mathcal L}=(N/M)a$ commensurate with the lattice period $a$, where $N$,$M$ are relatively prime numbers: (i) the propagated field $\psi(x,z)$ is periodic in $x$ with period $L=M \mathcal{L}=Na$; (ii) for any given small parameter $\epsilon$, there exists at least one propagation distance $z_0$ such that $\Delta(z_0)< \epsilon$, where 
\begin{equation}
\Delta(z)= (1/L) \int_{0}^{L}dx |\psi(x,z)-\psi(x,0)|^2
\end{equation}
measures the deviation of the propagated field, at distance $z$, from the initial field at $z=0$.
 \par
Theorem III states that in Hermitian crystals, even though Talbot revivals are unlikely, any periodic field distribution commensurate with the lattice period approximately reproduces itself with an arbitrary degree of accuracy after some (possibly long) propagation distance. Such a result is basically related to the quantum recurrence theorem of Hamiltonians with a discrete energy spectrum \cite{Loinger}.
 \begin{figure}
\includegraphics[scale=0.36]{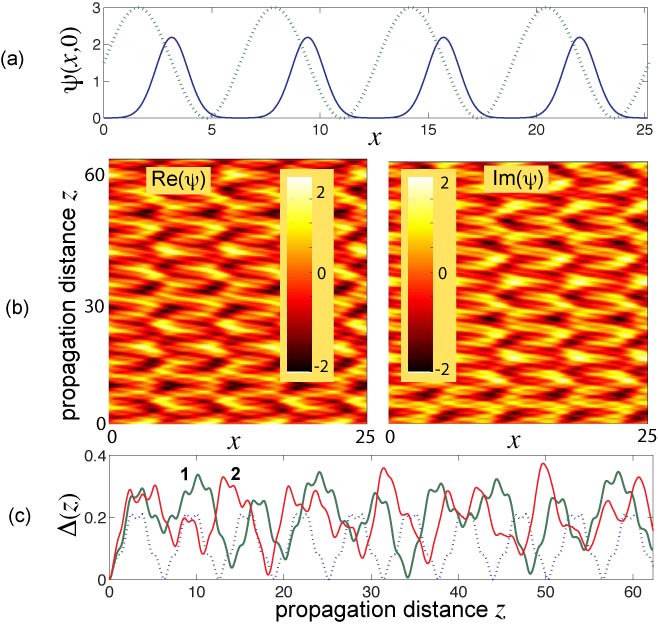}
\caption{(Color online) Quantum recurrence in the Hermitian Mathieu potential $V(x)=V_0 \sin(2 \pi x/a)$ for $a=2 \pi$. (a) Solid curve: input field distribution with spatial period $\mathcal{L}=a$; dotted curve: the Mathieu potential $V(x)$. (b) Maps of field propagation [real and imaginary parts of $\psi(x,z)$] for $V_0=2$. (c) Behavior of the the deviation function $\Delta(z)$, defined by Eq.(3), for  $V_0=1$ (curve 1) and $V_0=2$ (curve 2). The dotted curve in (c) corresponds to $V_0=0$ (free space propagation), with a revival period $z_T=2 \pi$.}
\end{figure}
\begin{figure}
\includegraphics[scale=0.36]{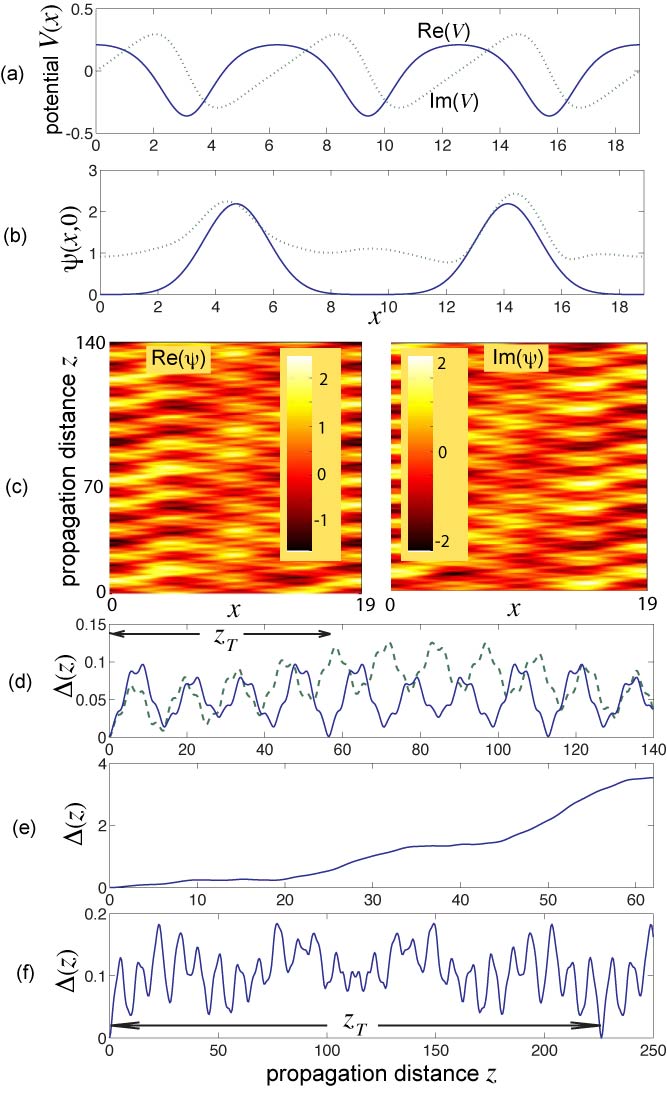}
\caption{(Color online) Talbot self-imaging in the $\mathcal{PT}$ crystal defined by Eq.(4). (a) Real and imaginary parts of the potential ($a=2 \pi$, $\rho=1$). (b) Initial field distribution (solid curve) with commensurate period $\mathcal{L}=(3/2)a$. The Talbot revival period is $z_T= (Na)^2/(2 \pi)\simeq 56.55$. The dotted curve shows the field distribution (in modulus) after a propagation distance $z=z_T/2$. (c) Maps of field propagation [real and imaginary parts of $\psi(x,z)$]. (d) Behavior of the deviation function $\Delta(z)$ (solid curve). The dashed curve in the figure shows the behavior of the deviation function corresponding to propagation in the associated Hermitian lattice, obtained by considering the real part solely of the potential (4). (e) Same as (d), but for an initial field distribution with period $\mathcal{L}=2a$. (f) Same as (e), but for a tilted field with the additional phase term $\exp(2 \pi i p x/a)$  ($p=1/3$). Self imaging is restored at the period $z_T= N^2a^2/(2 \pi p^2) \simeq 226.2$ according to theorem III.}
\end{figure}
An example of quantum recurrence is shown in Fig.2 for the Mathieu potential $V(x)=V_0 \sin( 2 \pi x /a)$.
The figure shows the propagation of a periodic waveform, composed by a sequence of Gaussians,  of period $\mathcal{L}=a$. The behavior of the deviation function $\Delta(z)$ [Fig.2(c)] clearly indicates that, even though self-imaging is not realized, there are propagation distances at which $\Delta(z)$ gets close to zero. 

\section{Talbot self-imaging in $\mathcal{PT}$-symmetric complex crystals}  A first example of gapless crystal is given by $V(x)=V_0 \exp(2 \pi i x/a)$, which has been considered in several previous studies \cite{Bragg2,Bender1,Longhi,Jones}. However, this potential shows a countable number of spectral singularities at energies $E= (\pi n/a)^2$ ($n=1,2,3,...$) \cite{Longhi,Jones}, which generally prevent the observation of self-images owing to the appearance of secularly-growing terms. The impact  of spectral singularities on self-imaging is discussed rather generally in Appendix B. Gapless crystals isospectral to the free space Hamiltonian, i.e. possessing the band structure of Fig.1(b) and that do not show spectral singularities at $q=0$, can be synthesized by supersymmetric (Darboux) transformations. Supersymmetric potentials with a given number (one, two, three, etc. ) of spectral singularities  can be constructed in this way \cite{Correra}. The simplest example is provided by the isospectral potential with a single spectral singularity at the energy $E=E_1=(\pi/a)^2$, which is given by (see Appendix C for technical details)
\begin{equation}
V(x)=\frac{(2 \pi / a)^2}{1+\cos \left( 2 \pi x/a+2 i \rho \right)}
\end{equation}
where $\rho$ is an arbitrary non-vanishing real number. This potential is $\mathcal{PT}$ symmetric [see Fig.3(a)] and the spectral singularity at $E=E_1$ indicates that we are at the symmetry breaking threshold. An example of self-imaging for this potential is shown in Fig.3, where the numerically-computed evolution of a periodic Gaussian train [Fig.3(b)] with $N/M=3/2$ is depicted. The onset of self-imaging is clearly observed by an inspection of Fig.3(d). The figure shows the behavior of the function $\Delta(z)$, defined by Eq.(3), which measures the deviation of the propagated field from its initial distribution. Vanishing of $\Delta(z)$ is the signature of self-imaging. The solid curve in Fig.3(d) depicts the behavior of $\Delta(z)$ corresponding to the potential (4). Note that, at exact revival distances $z=z_T, 2 z_T,...$, $\Delta(z)$ vanishes in this case. For comparison, the dashed curve in the same figure shows the behavior of $\Delta(x)$ for the Hermitian potential obtained by taking the real part of Eq.(4) solely. In this case self imaging is clearly not observed, indicating that the imaginary (non-conservative) part of the potential is necessary to achieve Talbot self images. It should be finally noticed that, contrary to the Talbot effect in free space, spatially-shifted replica of the initial field distribution is not observed for the $\mathcal{PT}$ symmetric potential (4) at odd multiplies of $z_T/2$. This is shown  in Fig.3(b), where the dotted curve depicts the field distribution of $\psi(x,z)$ (in modulus) at $z=z_T/2$.\\
According the general analysis presented in Sec.II,  a periodic field distribution which is commensurate with the lattice period but corresponding to an {\it even} value of $N$ is generally not self-imaged because of a secular growth arising from the spectral singularity at $E=E_1$. This is shown in Fig.3(e), where the evolution of $\Delta(z)$ is depicted for the propagation of the same Gaussian train of Fig.3(a), but with $N=2$ and $M=1$. Nevertheless, according to theorem III, tilting of the initial field distribution by an appropriate angle can avoid the secular growth and restores self-imaging. This is shown in Fig.3(f), where the self-imaging is restored for an initial Gaussian train distribution that excites the crystal at a suitable tilting angle.\par
\begin{figure}
\includegraphics[scale=0.36]{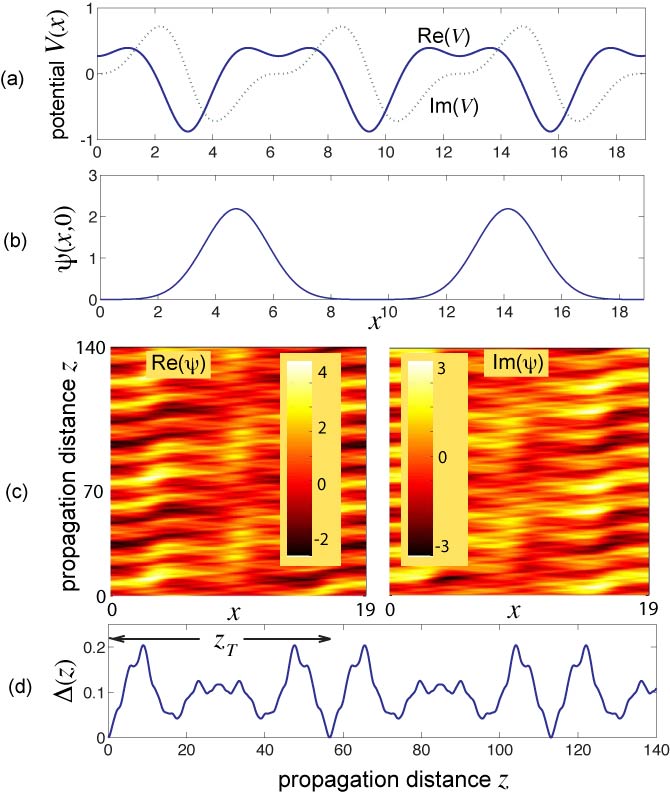}
\caption{(Color online) Talbot self-imaging in the $\mathcal{PT}$ crystal defined by Eq.(5). (a) Real and imaginary parts of the potential ($a=2 \pi$, $\rho=1$). (b) Initial field distribution (solid curve) with commensurate period $\mathcal{L}=(3/2)a$. The Talbot revival period is $z_T= (Na)^2/(2 \pi)\simeq 56.55$. (c) Maps of field propagation [real and imaginary parts of $\psi(x,z)$]. (d) Behavior of the deviation function $\Delta(z)$.}
\end{figure}
The gapless potential (4) is just one example of $\mathcal{PT}$-symmetric crystals that realize Talbot self-imaging. Other gapless potentials satisfying theorem I  can be synthesized by a cascading of supersymmetric transformations that introduce additional spectral singularities at energies $E=E_3=(3 \pi /a)^2$, $E=E_5=(5 \pi/ a)^2$, ..., but not at the energies $E_2$, $E_4$, ... For example, the complex crystal which is isospectral to the free-space Hamiltonian and that shows two spectral singularities at energies $E=E_1$ and $E=E_3$ is given by (see Appendix C for technical details) 
\begin{equation}
V(x)=\frac{(4 \pi /a)^2}{1-\cos \left( 4 \pi x/a+4 i \rho \right) } + \frac{ 2 (2 \pi / a)^2}{1+\cos \left( 2 \pi x/a+2 i \rho \right)}.
\end{equation}  
A typical behavior of the real and imaginary parts of the potential (5) is shown in Fig.4(a). Self-imaging in this potential is illustrated in panels (c) and (d) of Fig.4, which depict the evolution of the field [real and imaginary parts, panel (c)] and of the deviation function $\Delta(z)$ [panel (d)] for an initial periodic Gaussian train [panel (b)] and for $N=3$, $M=2$, $a= 2\pi$, corresponding to a Talbot distance $z_T= (Na)^2/(2 \pi)\simeq 56.55$.  Note that, at propagation distances $z=z_T, 2 z_T, ...$ the deviation function $\Delta(z)$ vanishes, indicating that exact self-imaging is realized. As compared to the potential (4), the addition of the second spectral singularity $E=E_3$ at the band edge $q= \pm \pi/a$ in the potential (5) does not introduce any special features in the dynamics, the Talbot distance $z_T$ being independent of the number of spectral singularities and being determined solely by the product $Na$.  

\section{Experimental implementation}
A challenging issue is the experimental realization of supersymmetric-generated complex potentials. Here we briefly mention that optical lattices with tailored gain and loss regions could provide experimentally accessible systems for the observation of commensurate Talbot self-images. Physical parameters of the lattices are similar to those estimated e.g. in Ref.\cite{Bragg2}. For example, for optical lattices realized in AlGaAs and probed in the near infrared ($\lambda=1.55 \; \mu$m, bulk refractive index $n_0 \simeq 3. 25$), for a lattice period of $20 \; \mu$m the Talbot revival distance $z_T$ in Fig.3(d) corresponds to a propagation length of $\simeq 15$ mm, whereas the maximum refractive index change  that realizes the potential of Fig.3(a) is $\Delta n_R \simeq 5.28 \times 10^{-4}$ and $\Delta n_I \simeq 2.77 \times 10^{-4}$ for the real and imaginary parts, respectively. While such values might be in principle feasible with the current technologies,  a precise and independent tailoring of the real and imaginary parts of the potential might be a challenging issue. Such a difficulty demands for the exploration  of other and more feasible experimental implementations  of the Schr\"{o}dinger equation with  a complex potential. Here we would like to briefly suggest   a fiber-optics implementation of the commensurate Talbot effect in a complex 'temporal' crystal. The basic idea is to consider the temporal evolution, at successive round trips, of an optical signal injected into  a dispersive fiber loop containing an amplitude and a phase modulator. Recirculating fiber loops have been since long time developed and investigated for long-haul transmission experiments in optical fiber communication systems and mode-locked lasers. A schematic of the dispersive fiber ring, containing a broadband optical amplifier, a phase modulator and an amplitude modulator, is depicted in Fig.5.
 \begin{figure}
\includegraphics[scale=0.45]{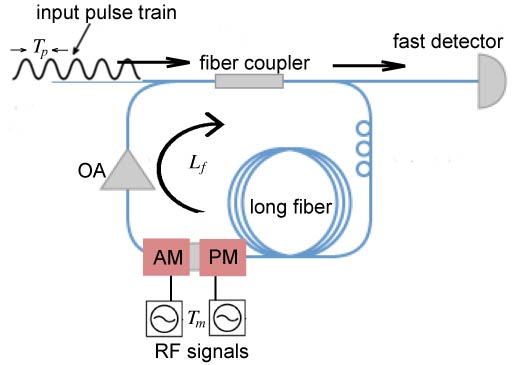}
\caption{(Color online) Schematic of a recirculating fiber loop that realizes Bragg scattering of an optical pulse train in a synthetic complex 'temporal' crystal. AM: amplitude modulator; PM: phase modulator; OA: broadband optical amplifier. The input pulse train comprises $N_p$ pulses spaced in time by $T_p$, which is commensurate with the modulation period $T_m$. $L_f$ is the total length of the ring. The evolution of the initial pulse train at successive round trips in the ring can be monitored at the output port of the fiber coupler using a fast photodetector. The amplitude and phase modulators are driven by two different but synchronized radio-frequency (RF) signals.}
\end{figure}
The ring is excited by a burst of $N_p$ optical pulses, spaced in time by $T_p$,  through the input port of an optical coupler (see Fig.5). The evolution of the optical pulse train at successive transits in the ring can be detected by monitoring the light signal at the output port of the coupler on a fast photodetector. The length $L_f$ of the fiber loop is long enough to accommodate the $N_p$ pulses of the injected train, avoiding overlapping (interference) effects at successive transits in the ring of delayed signals. The evolution at successive round trips of  the pulse train recirculating in the fiber loop can be described by the following master equation (see, for instance, \cite{master1,master2})
\begin{equation}
i \frac{\partial \psi}{\partial n}=- \mathcal{D} \frac{\partial^2 \psi}{\partial \tau^2}+ \left[ \delta_{PM}(\tau)+i \delta_{AM}(\tau) \right] \psi +i(g-l) \psi
\end{equation}
where $n$ is the round trip number, $\tau$ is the local time variable, $\psi(\tau,n)$ is the temporal envelope of the optical field in the ring at the $n$-th round trip, $\mathcal{D}$ is the total dispersion of the ring, $g$ and $l$ are  the optical gain coefficient and loss rate per transit, respectively, and $\delta_{PM}(\tau)$, $\delta_{AM}(\tau) $ are the periodic temporal profiles of phase and amplitude modulations impressed by the two modulators at each transit. Equation (6) holds provided that  the temporal waveform evolves little
during a single round trip in the ring and whenever nonlinear and finite bandwidth effects are negligible \cite{master2}. Note that Eq.(6) is a Schr\"{o}dinger-like equation with a complex potential, and thus it is suited to simulate the Bragg scattering problem in a complex crystal [Eq.(1)]. The temporal dispersion introduced by the fiber plays the same role as diffraction in free space, whereas the phase and amplitude changes impressed by the modulators mimic wave scattering off  the complex crystal. We note that, in the absence of the phase and amplitude modulators and for $g=l$, the dispersive wave equation (6) realizes in the temporal domain the analogue of the Talbot effect of paraxial diffraction theory in free space (temporal Talbot effect, see \cite{temporal}). It is worth introducing the scaled variables $x=\Omega_m \tau= 2 \pi \tau / T_m$ and $z=n \mathcal{D} \Omega_m^2$, where
 $T_m= 2 \pi/ \Omega_m$ is the modulation period of loss [$\delta_{AM}(\tau)$] and phase [$\delta_{PM}(\tau)$] introduced by the modulators. With such new variables, Eq.(6) assumes the form Eq.(1) with
\begin{equation}
V(x)=\frac{1}{\mathcal{D} \Omega_m^2} \left[ \delta_{PM}(x)+i \delta_{AM}(t)+i(g-l)\right] 
\end{equation}
and with $a= 2 \pi$. Note that in this way the real and imaginary parts of the complex potential $V(x)$ can be independently controlled and tailored by a suitable choice of the ac electrical signals driving the two modulators, making it possible to reproduce, for example, the potential defined by Eq.(4) \cite{note0}. The simplest case if the one where the two modulators are driven by the same sinusoidal signal, shifted by $\pi/4$ for AM and PM modulators, and with the same modulation depths: this realizes the complex sinusoidal optical potential  $V(x)=V_0 \exp( ix)$. The gain parameter  $g$ of the broadband optical amplifier is needed to compensate for cavity losses and to keep the cavity close to (but below) the lasing threshold. This enhances the lifetime of photons in the loop and enables to monitor the optical field at the output  port over several round trips. To observe self-imaging, the period $T_p$ of the injected pulse train should be commensurate with the modulation period $T_m$. Like in harmonically and actively mode-locked fiber ring lasers, active cavity length tuning is required to maintain a synchronous modulation over successive round-trips \cite{master2}. The total dispersion $\mathcal{D}$ of the loop is given by $\mathcal{D}=\lambda^2 D L_f /(4 \pi c)$, where $\lambda$ is the optical wavelength, $L_f$ is the fiber length, $D$ is the fiber dispersion, and $c$ is the speed o light in vacuum \cite{master2}. To get an idea of physical parameters corresponding to the commensurate Talbot effect shown in the simulations of Fig.3(c-d), let us consider a typical  fiber ring of length $L_f=100$ m, a modulation frequency $\nu_m=1/T_m=3$ GHz, and an optical pulse train at wavelength $\lambda=1560$ nm \cite{master2}.  To synthesize anharmonic waveforms like the ones shown in Fg.3(a), wide bandwidth modulators (e.g. 40 GHz bandwidth modulators, encompassing more than 10 harmonics of the fundamental harmonic) are needed. For fused silica fibers, the typical value of dispersion at $\lambda=1560$ nm is $D=50\; \rm { ps/(km \times nm)}$ \cite{master2}. The commensurate condition $N/M=3/2$ of Fig.3(c-d) requires a temporal spacing $T_p$ between adjacent pulses of the injected train equal to $T_p=(3/2)T_m \simeq 3.14$ ns. Note that the fiber loop is long enough to accommodate more than 100 pulses, avoiding aliasing effects \cite{note}. The Talbot self-imaging distance $z_T=18 \pi$ in Fig.3(d) corresponds to $n_T=z_T T_m^2/ (4 \pi^2 \mathcal{D}) \sim 4.9 \times 10^4$ round trips in the ring. From Eq.(7) one can also estimate the peak-to-peak modulation depths per round trip requested to realize the complex potential of Fig.3(a), which are of the order of  $\delta_{AM} \sim \delta_{PM} \sim 4 \pi^2 \mathcal{D} / T_m^2 \sim 0.002$. We note that the number of round trips in the cavity could be reduced, if needed, by increasing the modulation frequency $\nu_m$ or reducing $N$. For instance, at $\nu_m=6$ GHz and for $N=1$, Talbot self imaging requires  only $n_T \simeq 136 $ round trips.\par

\section{Conclusion and discussion} A conceptual extension of the Talbot effect in systems with discrete translational invariance has been introduced. The Talbot effect in crystals has been so far overlooked and regarded as a seldom phenomenon \cite{discTalbot}. Indeed, we showed rather generally that in ordinary crystals only approximate revivals are possible. The main novel finding here is that $\mathcal{PT}$-symmetric complex crystals, a new class of synthetic materials that are attracting a considerable attention recently \cite{uffa3,uffa4},  can show exact self imaging for input field distributions commensurate with the lattice period. Our results shed new light onto the physics of the Talbot effect in systems with discrete translation invariance, providing a link between an old phenomenon and a a new class of synthetic materials beyond the limitations found in Ref.\cite{PTTalbot}. On the experimental side, the realization of Bragg scattering in complex crystals is a rather challenging issue. In Sec.IV we have suggested an optical system where Bragg scattering occurs in a 'temporal' crystal, realized in a dispersive recirculating fiber loop with amplitude and phase modulators. However, different strategies could be investigated. On the theoretical side, one should finally mention that other important questions related to self-imaging phenomena in complex crystals have not been addressed in this work and should motivate further studies. For instance, can fractional or even fractal images be observed in any complex crystal? Can Talbot images with sub-wavelength resolution be obtained beyond the paraxial approximation in $\mathcal{PT}$ media?

 \appendix
 \section{Proofs of Theorems I, II and III}
 \subsection {Proof of Theorem I} 
Let us assume that $\psi(x,0)$ is periodic in $x$ with spatial period $\mathcal{L}$ which is commensurate with the lattice period $a$. This means that there exist two relatively prime integer numbers $N$ and $M$ such that $\mathcal{L}=(N/M)a$. Note that the case of an incommensurate period $\mathcal{L}$ can be obtained in the limit $N,M \rightarrow \infty$. The initial field distribution contains spatial harmonics with wave numbers $(2 \pi / \mathcal{L})l$ ($l=0, \pm 1, \pm2 , ...) $. 
Owing to Bragg scattering in the crystal, wave propagation introduces the additional spatial harmonics $(2 \pi /a)n$ ($n=0, \pm 1, \pm 2,...$), so that the propagated field $\psi(x,z)$ 
will be composed by the spatial harmonics 
\begin{equation}
\frac{2 \pi}{\mathcal{L}}l+\frac{2 \pi}{a}n=\frac{2 \pi}{L} \left( Ml+n N \right)
\end{equation}
where we have set
\begin{equation}
L=Na=M \mathcal{L}.
\end{equation}
This means that the propagated wave $\psi(x,z)$ will be a periodic function in $x$ with spatial period $L$, i.e. one can write
\begin{equation}
\psi(x,z)=\sum_{n=-\infty}^{\infty} \psi_n(z) \exp(2 \pi i n x / L)
\end{equation}
with 
\begin{equation}
\psi_n(0)=\frac{1}{L}\int_{0}^{L} dx \psi(x,0) \exp(- 2 \pi i n x /L).
\end{equation}
Note that, since $\psi(x,0)$ is periodic with period $\mathcal{L}=L/M$, one has $\psi_n(0)=0$ for $n \neq M l$ ($l=0, \pm 1, \pm 2 , ...$). The evolution of the amplitudes 
$\psi_n(z)$ is readily obtained after substitution of the Ansatz (A3) into the Schr\"{o}dinger equation $ i (\partial \psi / \partial z)=\hat{H} \psi$, where $\hat{H}=-\partial^2_x+V(x)$ [Eq.(1) in the text]. The analysis is simplified after introduction of the $N$ set of amplitudes $f_n(z;n_0) \equiv \psi_{n_0+n N}(z)$, where $n=0,\pm1,\pm2,\pm3,...$ and $n_0$ spans $N$ consecutive integer numbers, namely $n_0=-N/2,-N/2+1,...,0,1,...,N/2-1$ for $N$ even or $n_0=-(N-1)/2,-(N-1)/2+1,...,0,1,..., (N-1)/2$ for $N$ odd. One then obtains the following set of $N$ decoupled linear systems
\begin{equation}
i \frac{df_n}{dz}= \sum_{l=-\infty}^{\infty} \mathcal{H}_{n,l}(q_{n_0}) f_n(z;n_0)
\end{equation}
 where
\begin{equation}
\mathcal{H}_{n,l}(q_{n_0}) \equiv \left( q_{n_0} +\frac{2 \pi }{a}n \right)^2 \delta_{n,l}+V_{n-l},
\end{equation}
$q_{n_0} \equiv (2 \pi /a)(n_0/N)$, and  
\begin{equation}
V_n=\frac{1}{a} \int_{0}^{a}dx V(x)  \exp(-2 \pi i n a/x)
\end{equation}
 are the Fourier expansion amplitudes of the periodic potential $V(x)$. The matrix $\mathcal{H}(q_{n_0})=\mathcal{H}_{n,l}(q_{n_0})$ defined by Eq.(A-6) determines the energy spectrum and corresponding Bloch-Floquet eigenfunctions of the Hamiltonian $\hat{H}$ at the Bloch wave number $q=q_{n_0}$ (see, for instance, \cite{Longhi}). In particular, the eigenvalues of $\mathcal{H}(q_{n_0})$ are given by $E_{\alpha}(q_{n_0})$, where $\alpha=0,1,2,3,...$ is the band index \cite{Longhi}. For $q_{n_0} \neq -\pi/a,0$ the eigenvalues $E_{\alpha}(q_{n_0})$ are distinct and the solution to Eqs.(A5) is given by a linear superposition of terms oscillating like $\exp[-iE_{\alpha}(q_{n_0})z]$. At $q_{n_0}=0$ or $q_{n_0}=-\pi/a$, i.e. at the center or at  the edge of the Brillouin zone, eigenvalue degeneracy  arising from band touching (i.e. absence of a band gap) can arise. For an Hermitian crystal, i.e. for a real potential $V(x)$, such degeneracies do not correspond to defective eigenvalues (exceptional points), i.e. the Jordan matrix associated to $\mathcal{H}$ is diagonal: in this case the  solution to Eqs.(A5) is given again by a linear superposition of terms oscillating like $\exp[-iE_{\alpha}(q_{n_0})z]$, with $q_{n_0}=0, -\pi/a$. Hence, the spatial harmonic components $\psi_n(z)$ of the propagated field $\psi(x,z)$ [see Eq.(A3)] is a linear superposition of terms oscillating at the frequencies $\omega_{\alpha,n_0}=E_{\alpha}(q_{n_0})$, with $\alpha=0,1,2,3,....$ and 
 \begin{equation}
 q_{n_0}=-\frac{\pi}{a}, -\frac{\pi}{a}+ \frac{2 \pi}{Na}, ..., 0, \frac{2 \pi}{Na},...,\frac{\pi}{a}-\frac{2 \pi}{Na}
 \end{equation}
  for $N$ even, or 
  \begin{equation}
  q_{n_0}=-\frac{(N-1)\pi}{Na}, -\frac{(N-3)\pi}{Na} ..., 0,  \frac{2 \pi}{Na}, ... ,\frac{(N-1) \pi}{Na}
  \end{equation}
   for $N$ odd. Self-imaging at the propagation distance $z=z_T$ is obtained provided the numbers $\omega_{\alpha,n_0} z_T$ are integer multiplies than $ 2 \pi$. Such a condition is generally not satisfied in an Hermitian crystal because of the deviations of the band  dispersion relation from the free-space parabolic law. Instead, let us consider a complex crystal, i.e. let us assume that $V(x)$ has a non-vanishing imaginary part, and let us assume that (i) the energy spectrum of $\hat{H}$ is real and gapless, i.e. it is given by the semi-infinite line $E \in [0, \infty)$, and (ii) the band dispersion relations $E_\alpha(q)$ map the parabolic dispersion law of free space propagation [see Fig.1(b) in the text] 
   \begin{equation}
   E_{\alpha}(q)=(2 \pi \beta_\alpha / a -|q|)^2
   \end{equation}
   where $\beta_{\alpha}=0,1,-1,2,-2,3,-3,...$ for $\alpha=0,1,2,3,4,5,6,...$.
   In this case from Eqs.(A8), (A9) and (A10)  it readily follows that $\omega_{\alpha,n_0} z_T$ is an integer multiple of $2 \pi$ by assuming
 \begin{equation}
 z_T=\frac{N^2 a^2}{2 \pi}=\frac{M^2 \mathcal{L}^2}{2 \pi}
 \end{equation}
 which provides the spatial period of self-imaging. 
 However, some care should be paid owing to the appearance of spectral singularities in complex crystals. As shown in Refs.\cite{Longhi,Jones}, spectral singularities can appear at $q=-\pi/a$ or at $q=0$ and they correspond to defective eigenvalues (exceptional points) of the matrix $\mathcal{H}(q)$. In this case the Jordan form of $\mathcal{H}$ is no more diagonal and the solutions $f_n(z;n_0)$ for rather arbitrary initial conditions may show secularly growing terms in $z$, which prevent self-imaging (see also Appendix B). We note that for an even value of $N$ both values $q=-\pi/q$ and $q=0$ are included in the set $q_{n_0}$ [see Eq.(A8)], whereas for an odd value of $N$ only the value $q=0$ is included in the set $q_{n_0}$ [see Eq.(A9)]. Therefore, for an even value of $N$ self-imaging is possible provided that the spectrum of $\mathcal{H}$ does not have spectral singularities. Since the gapless and parabolic dispersion of the complex crystal is expected to be realizable at the $\mathcal{PT}$ symmetry breaking point, such a case has to be excluded. On the other hand, for an odd value  of $N$ it is sufficient to exclude the existence of spectral singularities at $q=0$, i.e. at the energies $E=(2 \pi l /a)^2$ ($l=0,1,2,3,...)$. This proves Theorem I. 
  \subsection {Proof of Theorem II} 
The proof of Theorem II follows the same lines stated in Theorem I. The propagated field can be expanded according to the Fourier decomposition [compare with Eq.(A3)]
 \begin{equation}
 \psi(x,z)=\sum_{n= - \infty}^{\infty} \psi_n(z) \exp (2 \pi i n x/L+ 2 \pi i p  x /a)
 \end{equation}
  where $L=Na$, and the evolution equations for $\psi_n$ can be derived, which take similar form as Eqs.(A5,A6).  Note that the propagated field contains the spatial harmonics  $k_n=2 \pi n/L+ 2 \pi p/a$ ($n=0, \pm 1, \pm 2, \pm3,..$), so that the energies of excited Bloch-Floquet states are now given by $E_n=k_n^2=(2 \pi  / a)^2 (p^2/N^2)(N+n/p)^2$. To realize self-imaging at propagation distances which are integer multiplies than $z_T$ the following two conditions should be met: (i) $E_nz_T$ should be integer multiplies than $ 2 \pi$ for any $n$, and (ii) $E_n$ should not correspond to band energies at $q=0$ or $q =-\pi/a$ (this is to avoid secular growths arising from spectral singularities). The former condition is satisfied by assuming $z_T=a^2N^2/(2 \pi p^2)$, whereas the latter condition requires that for any two arbitrary integers $n,l$ one has $2 (n+Np) \neq Nl$. Such a condition is safely  met whenever $2Np$ is not an integer number. This proves theorem II. Note that the initial field distribution $\psi(x,0)=f(x) \exp(2 \pi i p x/a)$ with $f(x+\mathcal{L})=f(x)$, $\mathcal{L}=(N/M)a$ is a subset of the periodic functions with period $Na/p$, and this explains the different expression for the self-imaging distance $z_T$ in theorem II as compared to Eq.(A11) in theorem I. 
  
\subsection {Proof of Theorem III} 
Theorem III is similar to the quantum recurrence theorem of Hamiltonians with a discrete energy spectrum \cite{Loinger} and its proof can be done following a similar procedure. Even though the spectrum of $\hat{H}$ is absolutely continuous and the quantum recurrence theorem does not generally hold in this case \cite{Loinger}, for the set of initial states which are periodic and commensurate with the lattice period the crystal basically behaves like an Hamiltonian system with  a discrete energy spectrum.
 As shown in the proof of Theorem I, the solution $\psi(x,z)$ to Eq.(1) corresponding to an initial field distribution $\psi(x,0)$ periodic with a period $\mathcal{L}=(N/M)a$ commensurate with the lattice period $a$ is periodic with period $L=Na$ and can be thus written as
 \begin{equation}
 \psi(x,z)=\sum_{n_0} \sum_{l=- \infty}^{\infty} f_l(z; n_0) \exp \left[ 2 \pi i (Nl+n_0)x/L \right]
 \end{equation}
 where $f_l(z;n_0)$ is the solution to the linear system (A5) and $n_0$ assumes $N$ consecutive integer values, as previously discussed. 
 Hence one has
 \begin{equation}
 \Delta (z)  \equiv \frac{1}{L}\int_{0}^{L}dx | \psi(x,z)-\psi(x,0)|^2=\sum_{n_0} R_{n_0}(z)
 \end{equation}
 where we have set
 \begin{equation}
 R_{n_0} (z) \equiv \sum_{l=-\infty}^{\infty} |f_{l}(z;n_0)-f_l(0;n_0)|^2.
 \end{equation}
 Note that, since $\hat{H}$ is Hermitian, $(1/L) \int_{0}^{L}dx |\psi(x,z)|^2$ is independent of $z$ and can be assumed equal to one by a suitable normalization of the initial field distribution. We wish to show that, for a fixed and arbitrarily small value of $\epsilon$, there exists a propagation distance $z=z_0$, depending on $\epsilon$ and possibly long, such that $\Delta(z_0) < \epsilon$, i.e. the field distribution $\psi(x,z)$ during propagation reproduces the initial one with any degree of accuracy after some propagation length $z=z_0$. To this aim, let us diagonalize the Hermitian matrix $\mathcal{H}(q_{n_0})$ entering in Eq.(A5) by setting $\mathcal{H}(q_{n_0})=\mathcal{T}^{-1}(q_{n_0}) \Lambda \mathcal{T}(q_{n_0})$, where $\Lambda$ is the diagonal matrix of the eigenvalues $E_{\alpha}(q_{n_0})$ and $\mathcal{T}(q_{n_0})$ is the unitary matrix of corresponding eigenvectors, i.e. $\mathcal{T}^{-1}(q_{n_0})=\mathcal{T}^{\dag}(q_{n_0})$. Then it can be readily shown that
 \begin{equation}
 R_{n_0}(z)=\sum_{\alpha} 2 |w_{\alpha}|^2 \left[ 1-\cos \left( E_{\alpha}(q_{n_0}) z \right) \right]
 \end{equation}
where $\mathbf{w}=\mathcal{T}(q_{n_0}) \mathbf{f}(0)$ and where we have set $\mathbf{w}=(w_0,w_1.w_2,....)^T$ and $\mathbf{f}(0)=( ..., f_{-1}(0,;q_{n_0} ), f_0(0,;q_{n_0}), f_1(0,;q_{n_0}),...)^T$.  Since the series $\sum_{\alpha} |w_{\alpha}|^2$ is convergent, there exists a (possibly large) integer number $Q$ such that 
\begin{equation}
\sum_{\alpha=Q+1}^{\infty}|w_{\alpha}|^2 < \frac{\epsilon}{8N}.
\end{equation}  
 Moreover, for a well-known property of  almost-periodic functions \cite{Loinger} the exists a propagation distance $z=z_0$ such that
\begin{equation}
\sum_{\alpha=1}^{Q} \left[1-\cos \left( E_{\alpha} (q_{n_0}) z_0 \right) \right]  < \frac{\epsilon}{4N} {\rm max}_{\alpha} \left( |w_{\alpha} |^2  \right)
\end{equation}  
and thus
\begin{equation}
\sum_{\alpha=0}^{Q}|w_{\alpha}|^2  \left[1-\cos \left( E_{\alpha} (q_{n_0}) z_0 \right) \right]   < \frac{\epsilon}{2N}
\end{equation}
 From Eqs.(A15), (A16),  and (A18) it then follows $R_{n_0}(z=z_0) < \epsilon/N$, and thus from Eq.(A14) one has $\Delta(z=z_0) < \epsilon$, which proves Theorem III. 
 
 \section{Role of spectral singularities} 
 As discussed in the proof of Theorem I (see Appendix A), a necessary condition for the observation of self-imaging is the absence of spectral singularities of $\hat{H}$ at energies $E_{\alpha}(q=0)$. In fact, a spectral singularity generally introduces a secular growth in $z$ of the propagating wave, which prevents the observation of Talbot self images. An example of a secular growth arising from spectral singularities is shown in Fig.3(e). To clarify this point, let us assume that at some energy $\mathcal{E}=E_{\alpha}(q=0)$ the Hamiltonian $\hat{H}$ has a spectral singularity. This means that there exist two periodic functions $u(x)$ and $v(x)$, of the same periodicity of the lattice, such that $(\hat{H}-\mathcal{E})u(x)= 0 $ and $(\hat{H}-\mathcal{E})v(x)=u(x)$. The functions $v(x)$ is the Jordan associated function to $u(x)$ \cite{Jones}. It can be readily shown that the Schr\"{o}dinger equation (1) with the initial condition $\psi(x,0)=v(x)$ is satisfied by taking
 \begin{equation}
 \psi(x,z)=-iz u(x) \exp(-i \mathcal{E}z)+v(x) \exp(-i \mathcal{E}z)
 \end{equation}
 which shows a secular growth with $z$. Hence an initial periodic field like $v(x)$ is not self-imaged when propagating in the crystal. This also holds for any other initial field distribution which is not "orthogonal" to $v(x)$. A sufficient condition to avoid a secular growth arising from spectral singularities is stated in Theorem II.
 \section{Synthesis of gapless complex crystals with a finite number of spectral singularities}
 Gapless crystals isospectral to the free space Hamiltonian possessing a given number (one, two, three, etc.) of spectral singularities  can be synthesized by supersymmetric (SUSY) transformations (see, for instance, \cite{Correra}). In particular, gapless crystals that do not show spectral singularities at $q=0$ can be obtained by a cascade of SUSY transformations of the free-particle Hamiltonian. For the sake of completeness, in this Appendix we derive the expressions Eqs.(4) and (5) of the isospectral gapless potentials introduced in Sec.III possessing one and two spectral singularities at energies $E_1= (\pi/a)^2$ and $E_3=(3 \pi/a)^2$.\\
 Let us first briefly review, for the sake of clearness, the SUSY machinery. Let us indicate by $\hat{H}_1=-\partial_x^2+V_1(x)$  the Hamiltonian corresponding to the potential $V_1(x)$, and let $\phi_1(x)$ be a solution (not necessarily normalizable) to the equation $\hat{H}_{1} \phi_1=E_1 \phi_1$. The Hamiltonian $\hat{H}_1$ can be then factorized as $\hat{H}_1=\hat{B}_1 \hat{A}_1 +E_1$, where $\hat{A}_1=-\partial_x+W_1(x)$,  $\hat{B}_1=\partial_x+W_1(x)$, and 
\begin{equation}
W_1(x)=\frac{(d \phi_1 / dx)}{\phi_1(x)}
\end{equation}
is the so-called super potential. The Hamiltonian $\hat{H}_2=\hat{A}_1 \hat{B}_1+E_1$, obtained by intertwining the operators ${\hat A}_1$ and $\hat{B}_1$, is called the partner Hamiltonian of $\hat{H}_1$. The following properties then hold:\\
(i) The potential $V_2(x)$ of the partner Hamiltonian $\hat{H}_2$ is given by
\begin{equation}
V_2(x)=W_1^2(x)-\frac{dW_1}{dx}+E_1=-V_1(x)+2E_1+2W_1^2(x)
\end{equation}
(ii) If $\hat{H}_1 \psi=E \psi$ with $E \neq E_1$, then $\hat{H}_2 \xi=E \xi$ with
\begin{equation}
\xi(x)=\hat{A}_1 \psi(x)=-\frac{d \psi}{dx}+W_1(x) \psi(x).
\end{equation}
(iii) The two linearly-independent solutions to the equation $\hat{H}_2 \xi= E \xi$ with energy $E=E_1$ are given by
\begin{equation}
\xi_1(x)=\frac{1}{\phi_1(x)} \; , \; \xi_2(x)= \frac{1}{\phi_1(x)} \int_0^x dt \phi_1^2(t).
\end{equation}
Let us apply the SUSY transformation by assuming $V_1(x)=0$ (free-particle Hamiltonian) and the following solution $\phi_1(x)$ to the equation $\hat{H}_1 \phi_1 =E_1 \phi_1$
\begin{equation}
\phi_1(x)= \cos (k_0x + i \rho)
\end{equation}
where $\rho, k_0$ are real  and positive parameters and $E_1=k_0^2$. As it will be shown below, the parameter $a= \pi / k_0$ will define the periodicity of the complex crystal isospectral to the free-particle Hamiltonian, whereas $E=E_1$ is a spectral singularity of $\hat{H}_2$. In fact, according to Eq.(C1) the super potential $W_1(x)$ is given by $W_1(x)=-k_0 {\rm tan} (k_0x+i \rho)$, so that from Eq.(C2) the potential $V_2(x)$ of the partner Hamiltonian $\hat{H}_2$ reads explicitly
\begin{equation}
V_2(x)=\frac{2 k_0^2}{ \cos^2 (k_0+i \rho)}=\frac{(2 \pi /a)^2 }{1+ \cos(2 \pi x/a+2 i \rho)} 
\end{equation}
which coincides with Eq.(4) given in the text. Note that $V_2(x)$ is a complex periodic potential, and -as previously anticipated- $a= \pi/k_0$ corresponds to the lattice period.  This means that the energy $E_1= (\pi/a)^2$ is reached at the band edge $q= \pm \pi/a$ of the Brillouin zone. Note that a non-vanishing value of $\rho$ is needed to avoid divergences of $V_2(x)$, making the potential effectively non-Hermitian and $\mathcal{PT}$ symmetric. Owing to the property (ii) stated above, the partner Hamiltonians $\hat{H}_1$ and $\hat{H}_{2}$ are isospectral, except for the energy $E=E_1$ which should be investigated separately. It can be shown (see \cite{Correra} for mode details) that $E=E_1$ is a "defective" eigenvalue, i.e. it corresponds to a spectral singularity. This follows from the fact that the two eigenfunctions $\xi_{\pm}(x)$ of $\hat{H}_2$ with energy $E$, obtained from Eq.(C3) by taking $\psi=\psi_{\pm}= \exp( \pm ikx)$ ($k= \sqrt{E}$), coalesce as $E \rightarrow E_1$. 
The energy $E=E_1$ is the {\it only} spectral singularity of $\hat{H}_2$, since for $E>0$, $E \neq E_1$ $\hat{H}_2$ has two linearly-independent  improper eigenfunctions.\\
Cascading of SUSY transformations can be used to synthesize gapless complex crystals with an arbitrary finite number of spectral singularities at $E=E_1,E_3,E_5,...$. Such potentials avoid spectral singularities at the band center $q=0$, i.e. at energies $E_{2}, E_4, E_6, ...$  [see Fig.1(b)], and thus satisfy the conditions of theorem I. As an example, let us synthesize the gapless complex crystal isospectral to the free-particle Hamiltonian $\hat{H}_1$ and showing two spectral singularities at energies $E_1= (\pi/a)^2$ and $E_3=(3 \pi/a)^2$. To this aim, let us notice that a solution to the equation $\hat{H}_2 \phi_2=E_3 \phi_2$ is given by
\begin{equation}
\phi_2(x)=3 k_0 \sin(3k_0x+3 i \rho)-k_0 \tan (k_0x+i \rho) \cos(3 k_0x+3i \rho)
\end{equation}
To derive Eq.(C7), we used property (ii) of SUSY stated above and the trivial fact that $\cos(3k_0x+3i \rho)$ is an eigenfunction of $\hat{H}_1$ with energy $E_3$. We can then introduce the super potential $W_2(x)=(d \phi_2 /dx) / \phi_2(x)$ for $\hat{H}_2$, such that $\hat{H}_2=\hat{B}_2 \hat{A}_2+E_3$ with $\hat{A}_2=-\partial_x+W_2$, $\hat{B}_{2}=\partial_x+W_2$. 
From Eq.(C7), the super potential $W_2(x)$ can be readily calculated and reads explicitly
\begin{equation}
W_2(x)=k_0 \frac{3 \cos^2(k_0x+i \rho)-2}{\sin(k_0x+i \rho) \cos(k_0x+i \rho)}.
\end{equation}
The partner Hamiltonian $\hat{H}_3=\hat{A}_2 \hat{B}_2+E_3$, obtained from $\hat{H}_2$ by intertwining the operators $\hat{A}_2$ and $\hat{B}_2$, is associated with the potential
\begin{equation}
V_3(x)=-V_2(x)+2 W_2^2(x)+2E_3.
\end{equation}
Substitution of Eqs.(C6) and (C8) into Eq.(C9) yields
\begin{equation}
V_3(x)=\frac{(4 \pi /a)^2}{1-\cos \left( 4 \pi x/a+4 i \rho \right) } + \frac{ 2 (2 \pi / a)^2}{1+\cos \left( 2 \pi x/a+2 i \rho \right)}
\end{equation}   
 which is precisely Eq.(5) given in the text.


\begin{thebibliography}{}
\bibitem{note}
 The number of pulses $N_0$ needed to fill the entire ring can be estimated as $N_0 \sim T_R/T_p$, where $T_R=n_0 L_f/c$ is the transit time of light in the ring and $n_0$ is the 
 fiber refractive index. Taking $L_f=100$ m, $n_0 \simeq 1.47$ for fused silica at $\lambda=1560$ nm and $T_p \simeq 3.14$ ns, one has $N_0 \sim 156$. To avoid interference effects of pulse replica at successive transits in the loop, the number of pulses $N_p$ of the injected train should be smaller than $N_0$. For example, taking $N_p/N_0 \sim  2/3$, one has $N_p \sim 104$.
\end{thebibliography}
\end{document}